\documentclass{article}

\usepackage{arxiv}
\usepackage{xltabular}
\usepackage[utf8]{inputenc} 
\usepackage[T1]{fontenc}    
\usepackage{hyperref}       
\usepackage{url}            
\usepackage{booktabs}       
\usepackage[table]{xcolor} 
\usepackage{amsfonts}       
\usepackage{nicefrac}       
\usepackage{mathtools}      
\usepackage{microtype}      
\usepackage{lipsum}		
\usepackage{graphicx}
\usepackage[numbers]{natbib}
\usepackage{doi}
\usepackage{longtable}
\usepackage{titlesec}
\titlespacing*{\subsection}{0pt}{0.6ex}{0.4ex}
\titlespacing*{\subsubsection}{0pt}{0.2ex}{0pt}
\newcolumntype{L}[1]{>{\RaggedRight\arraybackslash}p{#1}}
\usepackage{booktabs,tabularx,array,ragged2e,float,placeins}
\usepackage{amsmath}

\title{VisitHGNN: Heterogeneous Graph Neural Networks for Modeling Point-of-Interest Visit Patterns}

\author{
  {\includegraphics[scale=0.06]{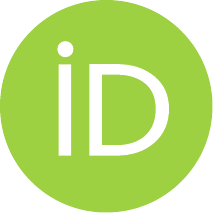}\hspace{1mm}Lin Pang} \\
  Graduate Research Assistant\\
  College of Engineering, University of Georgia\\
  Athens, GA, USA\\
  \texttt{Lin.Pang@uga.edu} \\
  \And
  {\includegraphics[scale=0.06]{orcid.pdf}\hspace{1mm}Jidong J.~Yang} \\
  Associate Professor\\
  College of Engineering, University of Georgia\\
  Athens, GA, USA\\
  \texttt{Jidong.Yang@uga.edu} \\
}

\date{October 2, 2025 }

\renewcommand{\shorttitle}{\textit{VisitHGNN}}

\hypersetup{
  pdftitle={VisitHGNN},
  pdfsubject={cs.LG, cs.SI, stat.ML}, 
  pdfauthor={Lin Pang, Jidong J.~Yang},
  pdfkeywords={Heterogeneous GNN, POI, Mobility, Representation Learning, Visit Prediction},
}

\begin{document}
\raggedbottom

\maketitle

\begin{abstract}
Understanding how urban residents travel between neighborhoods and destinations is critical for transportation planning, mobility management, and public health. By mining historical origin-to-destination flow patterns with spatial, temporal, and functional relations among urban places,  we estimate probabilities of visits from neighborhoods to specific destinations. These probabilities capture neighborhood-level contributions to citywide vehicular and foot traffic, supporting demand estimation, accessibility assessment, and multimodal planning. Particularly, we introduce VisitHGNN, a heterogeneous, relation-specific graph neural network designed to predict visit probabilities at individual Points of interest (POIs). POIs are characterized using numerical, JSON-derived, and textual attributes, augmented with fixed summaries of POI--POI spatial proximity, temporal co-activity, and brand affinity, while census block groups (CBGs) are described with 72 socio-demographic variables. CBGs are connected via spatial adjacency, and POIs and CBGs are linked through distance-annotated cross-type edges. Inference is constrained to a distance-based candidate set of plausible origin CBGs, and training minimizes a masked Kullback--Leibler (KL) divergence to yield probability distribution across the candidate set. Using weekly mobility data from Fulton County, Georgia, USA, VisitHGNN achieves strong predictive performance with mean KL divergence of \(0.287\), MAE of \(0.008\), Top-1 accuracy of \(0.853\), and \(R^{2}=0.892\), substantially outperforming pairwise MLP and distance-only baselines, and aligning closely with empirical visitation patterns (NDCG@50 \(= 0.966\); Recall@5 \(= 0.611\)). The resulting distributions closely mirror observed travel behavior with high fidelity, highlighting the model's potential for decision support in urban planning, transportation policy, mobility system design, and public health.

\end{abstract}


\section{Introduction}
Human mobility, characterized by the daily movement between neighborhoods and the places where people work, shop, and socialize, fundamentally shapes urban planning, transportation systems, retail siting, and public health.  Developing models that rigorously capture  mobility mechanisms is therefore of both scientific and practical importance. Decades of empirical research demonstrate that human movement is supported by rational decision: individuals repeatedly return to a small set of highly frequented locations with preferred distances of travel \cite{Gonzalez2008Understanding}, yielding reproducible aggregate patterns that, through entropy-based analysis, exhibit high theoretical predictability \cite{Song2010Limits}. When origin–destination flow data are integrated into computational models, they provide critical insights into the coupled dynamics among travel demand\cite{Rong2025SurveyOD}, infrastructure supply, and urban development, particularly at the level of POIs, where many planning and operational decisions are made\cite{Chang2021MobilityCOVID}. 

Despite advances from classical spatial-interaction models to modern machine learning and deep learning, the field still lacks a framework capable of producing POI-level, distributional, and well-calibrated visitor–origin predictions while explicitly modeling heterogeneous relations. Traditional models, such as gravity\cite{Jung2008GravityKoreanHighway}, intervening opportunities \cite{Intervening}, and radiation \cite{Simini2012UniversalModel}, provide interpretable, distance- and population-based priors, and recent variants incorporate social media, traffic counts, land use, or travel-cost/sociodemographic factors to improve model fitting \cite{Ren2014RadiationTemporal, Lenormand2015SocioDemoMobility}. Nonetheless, their fixed functional forms and reliance on hand-engineered components limit fine-scale predictive power and the ability to capture heterogeneous spatial interactions; consequently, attention has been shifted to nonlinear, data-driven methods. When enriched with socioeconomic and large-scale behavioral features, classic machine-learning approaches achieve higher accuracy than traditional analytical models, with kernel- and tree-based methods often surpassing conventional neural network baselines \cite{RodriguezRueda2021ODMatrixML,Simini2021DeepGravity,Robinson,Cai2022SpatialAttentionOD,Pourebrahim2019TripTwitter,PourebrahimEnhancingTwitter}. Nevertheless, they typically formulate visit prediction as independent tabular regression, which underutilized the intrinsic urban graph and its multi-relational structure (e.g., spatial proximity, functional similarity, temporal co-activity). Recent advances in deep learning increasingly recast visit prediction as spatial–relational representation learning problems. Grid-based CNN–GNN hybrids, for example, learn regional embeddings from multimodal signals to forecast OD volumes \cite{Cai2022SpatialAttentionOD}, but the gridding process blurs POI boundaries and imposes coarse cell-level assumptions that limit POI-level inference. Geo-contextual GAT models with multitask objectives capture spatial correlations and fuse learned embeddings with a boosted-tree prediction head \cite{Liu2020GeoContextualEmbeddings}; however they employ simple-relation, region-level graphs and non-probabilistic heads. Graph neural networks (GNN) applied to adjacency-based graphs improve imputation and prediction via neighborhood similarity and negative sampling \cite{Yao2021ODImputationGCN}, but typically rely on homogeneous graphs. Causal approaches aim for cross-city out-of-distribution robustness through factor disentanglement and sample re-weighting \cite{Zeng2022CausalOD}, yet they are generally assessed at regional scales and do not yield POI-level predictions. Meanwhile, the POI-embedding literature demonstrates that rich POI features, such as geographic, co-occurrence, and textual description, can be effectively encoded through multi-context sampling and attention mechanisms \cite{AdaptGOT}. Still, such embeddings are often task-specific and seldom integrated into heterogeneous POI--POI or POI--CBG graphs.

We hypothesize that POI-level visit prediction improves when rich POI features are fused with CBG socio-demographics through relation-aware POI-CBG links. Specifically, the POI nodes incorporate operating-hours, categorical, textual, and dwell-time features, augmented with fixed summaries of POI–POI spatial proximity, temporal co-activity, and brand affinity, which are treated as node attributes rather than message-passing edges. The CBG nodes contribute 72 socio-demographic and commuting indicators and are connected via spatial adjacency. Aggregation occurs only along distance-annotated POI--CBG candidate and belong links (indicating whether a POI belongs to a CBG), yielding more accurate origin distributions than models that collapse spatial detail or disregard relational heterogeneity\cite{wang2021,sch,vegra}.
We further posit that framing origin prediction as a probability distribution over a distance-based candidate set of plausible neighborhoods, and optimizing a masked KL divergence, improves reliability and ranking accuracy while enhancing computational efficiency compared to unconstrained or point-estimate objectives \cite{Simini2012UniversalModel,gu2009,guo2017}. 
Collectively, these design choices define our proposed VisitHGNN, a heterogeneous, multi-relational graph neural network that performs relation-aware fusion across CBG--POI  connections to generate decision-ready origin maps. The model consistently surpasses strong baselines across divergence, error, and ranking metrics. 

\begin{figure}
    \centering
    \includegraphics[width=1\linewidth]{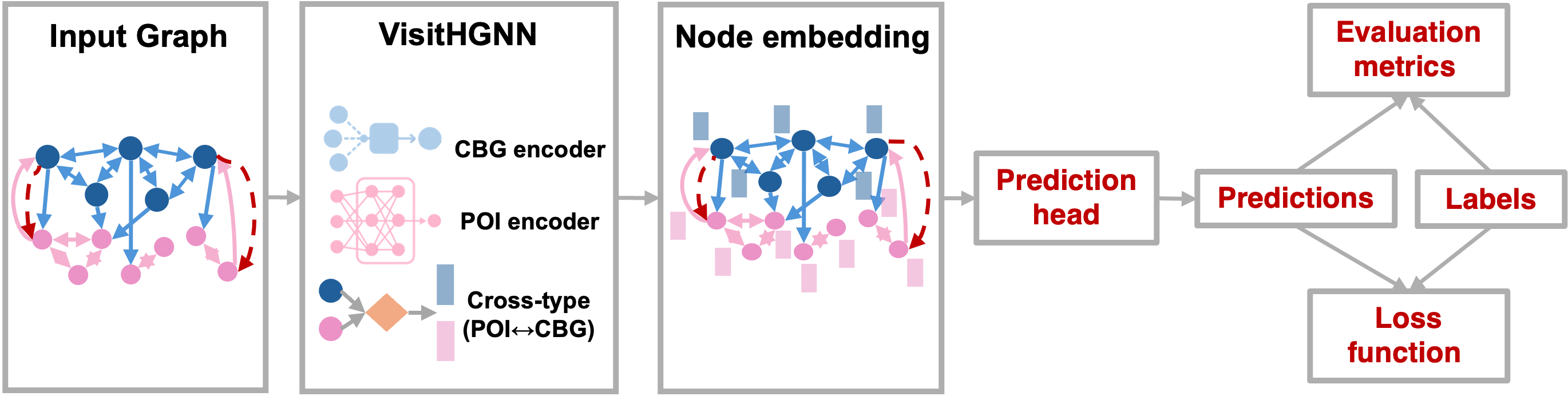}
    \caption{High-level pipeline of VisitHGNN}
    \label{Figureoverall_pipeline}
\end{figure}

As illustrated  in Figure~\ref{Figureoverall_pipeline}, the VisitHGNN workflow proceeds in four stages:

\begin{enumerate}
  \item Graph construction: build a transductive heterogeneous graph from weekly mobility data, incorporating CBG--CBG adjacency, POI--CBG \emph{belong} and distance-based KNN links, and POI--POI geospatial, temporal, and brand relations, where available.
  \item Node encoding: Generate initial node representations, with optional refinement of POI embeddings via relation-specific message passing.
  \item Cross-type fusion: Integrate relation-specific embeddings to obtain final node representations $(z_{\mathrm{CBG}}, z_{\mathrm{POI}})$.
  \item  Probabilistic inference: Apply a masked softmax  prediction head over each POI’s candidate set $N_{K}(u)$ to estimate origin distributions.  
\end{enumerate}


Training minimizes a masked KL divergence, while evaluation employs a transductive split in which validation and test POI visit labels are withheld during training. On weekly mobility data from Fulton County, Georgia, VisitHGNN achieves $\mathrm{KL}=0.287$, $\mathrm{MAE}=0.008$, Top\mbox{-}1 $=0.853$, and $R^{2}=0.892$, outperforming both a distance-decay baseline and a pairwise multilayer perceptron under matched supervision and candidate sets. Ablation studies further demonstrate the contribution of  POI--POI relational summaries, CBG adjacency, cross-type connectivity, and GraphNorm.

The remainder of this paper is organized as follows:
Section~\ref{sec:data} describes the dataset and split, Section~\ref{sec:method} introduces the VisitHGNN architecture, Section~\ref{sec:pred} details the prediction and training procedures, and Section~\ref{sec:experiments} presents experimental results and analysis.

\section{Data and Transductive Split}
\label{sec:data}
\begin{figure}[H]
    \centering
    \includegraphics[width=0.95\linewidth]{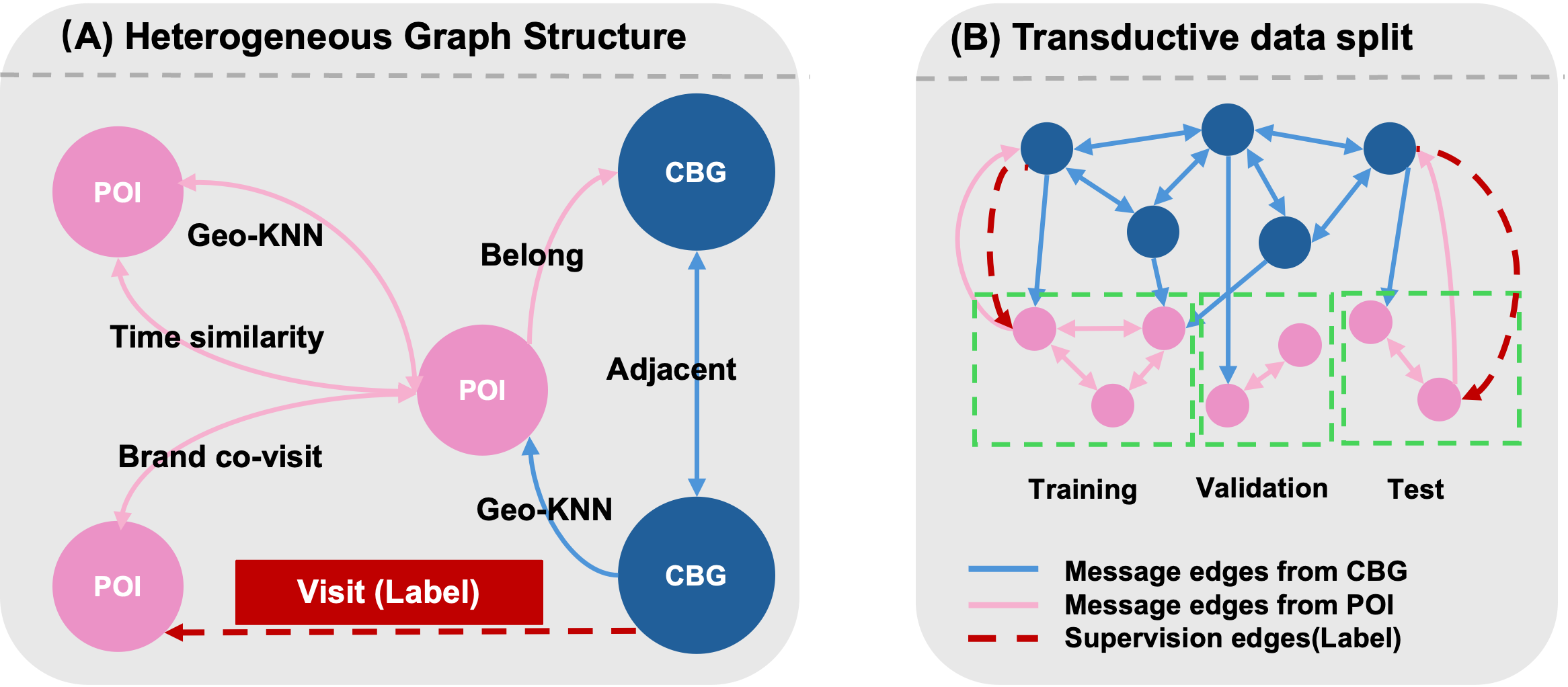}
    \caption{Heterogeneous graph overview and transductive data split.}
    \label{Figuregraphsplit}
\end{figure}

We represent weekly mobility patterns as a transductive heterogeneous graph composed of two node types: POIs and CBGs that are connected within a unified framework. The full message-passing topology is preserved across all data splits to maintain consistent relational context, while edges encoding visits are used exclusively as supervision signals.  As a case study, we employ data from Fulton County, Georgia.  POI nodes are partitioned into training, validation, and test sets, whereas some CBG nodes may be shared across splits depending on their relationships with POIs. Under the transductive setting, the model accesses the complete graph structure during training but not the ground-truth visit distributions of the validation or test POIs, enabling evaluation under realistic transductive conditions. The schema of node relationships and data partitioning strategy are illustrated in Figure~\ref{Figuregraphsplit}.

\subsection{Heterogeneous Graph Structure}
\label{subsec:data-entities}
The heterogeneous graph consists of two node types:POIs and CBGs, which are connected by a diverse set of edges specifically designed for learning proper representation. CBG–CBG edges encode geographic contiguity (“adjacent”), enforcing local spatial smoothness among neighborhoods. POI–CBG edges include a structural “belong” link mapping each POI to its home CBG and a distance-based K-nearest-neighbor (KNN) link connecting each POI to nearby CBGs to delineate plausible origin sets. POI–POI edges capture three dependencies: geospatial KNN reflecting inter-POI proximity, temporal similarity reflecting alignment of activity profiles, and a brand/co-visit relation, when available, capturing shared identity or empirically co-visited venues. Directed CBG-POI “visit” edges record observed flows and are held out as supervision targets rather than used for message passing. For clarity, relations originating from CBGs are colored blue and those originating from POIs are colored pink in the schematic (Fig. \ref{Figuregraphsplit}A).

\subsubsection{POI nodes and features}

Each POI node is encoded from three complementary raw formats: direct numeric fields, JSON objects, and text-based attributes. Missing entries are imputed as zero. Table~\ref{tab:poi_features} summarizes the complete feature set by source. Numeric features comprise polygon footprint area, total and unique visit counts, median distance from home, median dwell time, and provider-supplied normalized visit metrics. Two additional families of features are extracted from JSON objects: (1) Opening hours:  expanded into weekday-specific totals, with explicit handling of overnight spans, plus an aggregate weekly open–day count; and (2) Dwell-time distributions: bucketed durations mapped to seven fixed bins and \(\ell_1\)-normalized per POI. Text-derived attributes capture the provider’s top business category and the two-digit NAICS sector. Consistent codebooks are maintained across weeks to ensure stable indexing. Unless noted otherwise, these features constitute the baseline node representation. 
To enrich these representations, we embed standardized POI metadata, including venue name, brand, categories, region, and website tags, using a BERT-based language model to generate dense feature vectors \cite{devlin2019bert}. Additionally, normalized dwell-time histograms, derived from bucketed visit durations, are incorporated as complementary attributes. These embeddings and features are aligned and integrated with the baseline representation across all POI nodes in the graph.

{%
\setlength{\tabcolsep}{4pt}
\begin{xltabular}{\textwidth}{@{}L{0.25\textwidth} L{0.05\textwidth} L{0.27\textwidth} L{0.36\textwidth}@{}}
\caption{POI node features grouped by source format.}
\label{tab:poi_features}\\
\toprule
\textbf{Source column(s)} & \textbf{Dim.} & \textbf{Feature names} & \textbf{Description} \\
\midrule
\endfirsthead

\toprule
\textbf{Source column(s)} & \textbf{Dim.} & \textbf{Feature names} & \textbf{Description} \\
\midrule
\endhead

\bottomrule
\endfoot

\multicolumn{4}{@{}l}{\textbf{Raw numeric}} \\
WKT area (\(\mathrm{m}^{2}\)) & 1 & wkt\_area\_sq\_meters &
The calculated area of the polygon in \(\mathrm{m}^{2}\). \\
Raw visit counts & 1 & raw\_visit\_counts &
Total visits during the analysis window. \\
Raw visitor counts & 1 & raw\_visitor\_counts &
Unique panel visitors during the analysis window. \\
Distance from home (\(\mathrm{m}\)) & 1 & distance\_from\_home &
Median distance from visitors’ inferred home to the POI (meters). \\
Median dwell (\(\mathrm{min}\)) & 1 & median\_dwell &
Median visit duration (minutes). \\
Normalized visits & 5 &
state\_scaled, region\_visits, region\_visitors, total\_visits, total\_visitors &
Panel-normalized counts at state, region, and overall levels. \\

\midrule

\multicolumn{4}{@{}l}{\textbf{JSON-derived}} \\
Opening hours & 8 &
open\_mon\_hours \makebox[\linewidth][l]{\(\cdots\)} open\_sun\_hours; open\_days\_in\_week &
Daily operating hours by weekday (with overnight handling) and the weekly count of open days. \\
Bucketed dwell distribution & 7 & dwell\_time\_histogram &
Share of visits in seven dwell-time bins: \(<5\), \(5\text{–}10\), \(11\text{–}20\), \(21\text{–}60\), \(61\text{–}120\), \(121\text{–}240\), \(>240\) minutes; row-wise \(L_{1}\) normalized. \\

\midrule

\multicolumn{4}{@{}l}{\textbf{Text-derived}} \\
Category labels & 2 & top\_category, naics\_code &
Encodes the top business category and the two-digit NAICS sector using a stable codebook across weeks. \\
Aggregated text fields & 768 & text\_embedding &
Concatenates name, brand, categories and tags, city/region, and website into a single cleaned string, then embeds it with a pretrained BERT model; the \([{\rm CLS}]\) token vector is used. \\

\end{xltabular}
}

We construct a multi-relational graph of POIs that jointly captures geospatial, functional, and temporal associations. 

\begin{itemize}
  \item Geospatial channel: Each POI is linked to its \(K\) nearest neighbors using the haversine metric. Edges encode both great-circle distance (in kilometers) and compass bearing \(0\text{–}360^{\circ}\), enabling directional effects to be modeled alongside spatial separation. 
  \item Functional channel: Brand affinity is inferred from co-visitation patterns within a visitor’s origin CBG. Brand names are standardized, mapped to venues, and aggregated into weighted co-occurrence links. 
 \item Temporal channel: Co-activity is measured by similarity in hourly demand rhythms. Each 168-hour visit profile is \(\ell_1\)-normalized, and cosine similarity is used to connect the top-\(K\) most similar peers. 
\end{itemize}
 

This POI-POI graph representation provides a data-calibrated relational prior to enable learning of context-aware POI embeddings.

\subsubsection{CBG nodes and features}
Each CBG node is characterized by 72 socio–demographic variables from the 2019 American Community Survey \cite{census2019}, encompassing population composition, household economics, and commuting patterns. These variables include absolute counts and normalized percentages, summarized in Table~\ref{tab:acs_vars_summary}. Spatial anchors are derived from geographic centroids of CBG polygons from the 2019 GISshapefile. The final CBG representation integrates these socio-demographic attributes with centroid coordinates, yielding a feature set that reflects both the social profile and spatial context of each neighborhood.

\begin{table}[H]
\caption{Summary of CBG node features from ACS.}
\label{tab:acs_vars_summary}
\centering
\small
\setlength{\tabcolsep}{4pt}
\renewcommand{\arraystretch}{1.15}
\begin{tabularx}{\linewidth}{@{}L{0.30\linewidth} >{\RaggedRight\arraybackslash}X@{}}
\toprule
\textbf{Feature Category} & \textbf{Description} \\
\midrule
Population Structure & Population size, age groups, sex, and race. \\
Labor Market & Labor force participation and employment status for the population aged 16+. \\
Household Economics & Household income brackets and poverty indicators. \\
Educational Attainment & Education levels for population aged 25+. \\
Marital Status & Shares by marital status for population aged 15+. \\
Commuting Mode & Mode shares including private vehicle, public transit, walking, or biking. \\
Commuting Travel Time & Distribution of commute durations. \\
Commuting Departure Time & Departure times across morning, afternoon, and night. \\
Health Insurance & Coverage rates among the civilian non-institutionalized population. \\
\bottomrule
\end{tabularx}
\end{table}

\subsubsection{Relations among nodes}
The heterogeneous graph comprises two node types—POIs and CBGs—connected through relation-specific edges for representation learning:
\begin{enumerate}
  \item \textbf{CBG--CBG relations:} Geographic neighborhood structure is captured by unweighted ``adjacent'' edges, defined via topological adjacency among CBG polygons.

  \item \textbf{POI--CBG relations:} Two spatial linkages are considered: (1) unweighted ``belong'' edges connecting each POI to its administrative CBG; and (2) KNN edges linking POIs to their nearest CBG centroids, with Euclidean distance stored as an edge attribute. All message-passing edges are treated as undirected to enable bidirectional information flow; explicit self-loops are omitted. Self-features are handled by each layer's self-transform rather than by loop edges. Supervision is provided through directed \texttt{visit} edges from CBGs to POIs, which capture observed visitor flows. For each POI, \texttt{visit} edge weights are normalized to yield a probability distribution over origin CBGs. These labels are reserved for training and evaluation and do not participate in message passing.

  \item \textbf{POI--POI relations:} Three complementary edge types capture spatial, temporal, and functional coupling {\texttt{geo\_KNN},\texttt{time\_sim},\texttt{brand}}.
\end{enumerate}


Each POI–POI edge records its channel identifier and a unified attribute vector, enabling the model to jointly integrate these signals during message passing on the heterogeneous graph \cite{lei2025}.

\subsection{Transductive split and hyperparameter search}
\label{subsec:split}
The data partitioning and label masking strategy are depicted in Figure~\ref{Figuregraphsplit}B. Evaluation is performed in a transductive setting, where the model has full access to the entire graph topology during training, including node features and message-passing edges for all POIs and CBGs.  Given the target task being the prediction of POI visits, POI nodes are randomly partitioned into training (70\%), validation (15\%), and test (15\%) sets. Split indices are embedded in the graph object,  maintaining consistency across preprocessing, batching, and evaluation. All structural and relational edges span the full graph, allowing the GNN to aggregate neighborhood context for validation and test nodes without accessing their labels. Supervision edges connecting CBGs to POIs are partitioned according to the POI split. During training, the loss is computed only on visit edges linked to training POIs. The validation set is used exclusively for hyperparameter tuning, while the test set is reserved for final performance evaluation.

Hyperparameter optimization for VisitHGNN was conducted using the Optuna framework with a Tree-structured Parzen Estimator (TPE) sampler over a unified search space \cite{akiba2019optuna}. The model training excludes all visit edges from message passing, reserving them strictly for supervision, while retaining spatial and relational edges including POI--CBG KNN edges and CBG adjacency edges. When enabled, POI–POI channels {\texttt{geo\_KNN},\texttt{time\_sim},\texttt{brand}} and their associated edge attributes are also incorporated. To ensure fair comparisons, a fixed top-K prediction parameter was optionally enforced, enabling consistent candidate set sizes across experiments. Training and validation graphs were dynamically constructed to guarantee proper alignment of edge indices and attributes, with self-loops excluded. Early stopping based on validation KL divergence was employed to mitigate overfitting. The best-performing hyperparameters were retained for reproducibility and subsequent evaluation.


\section{The VisitHGNN Model}
\label{sec:method}
\begin{figure}[H]
    \centering
    \includegraphics[width=1\linewidth]{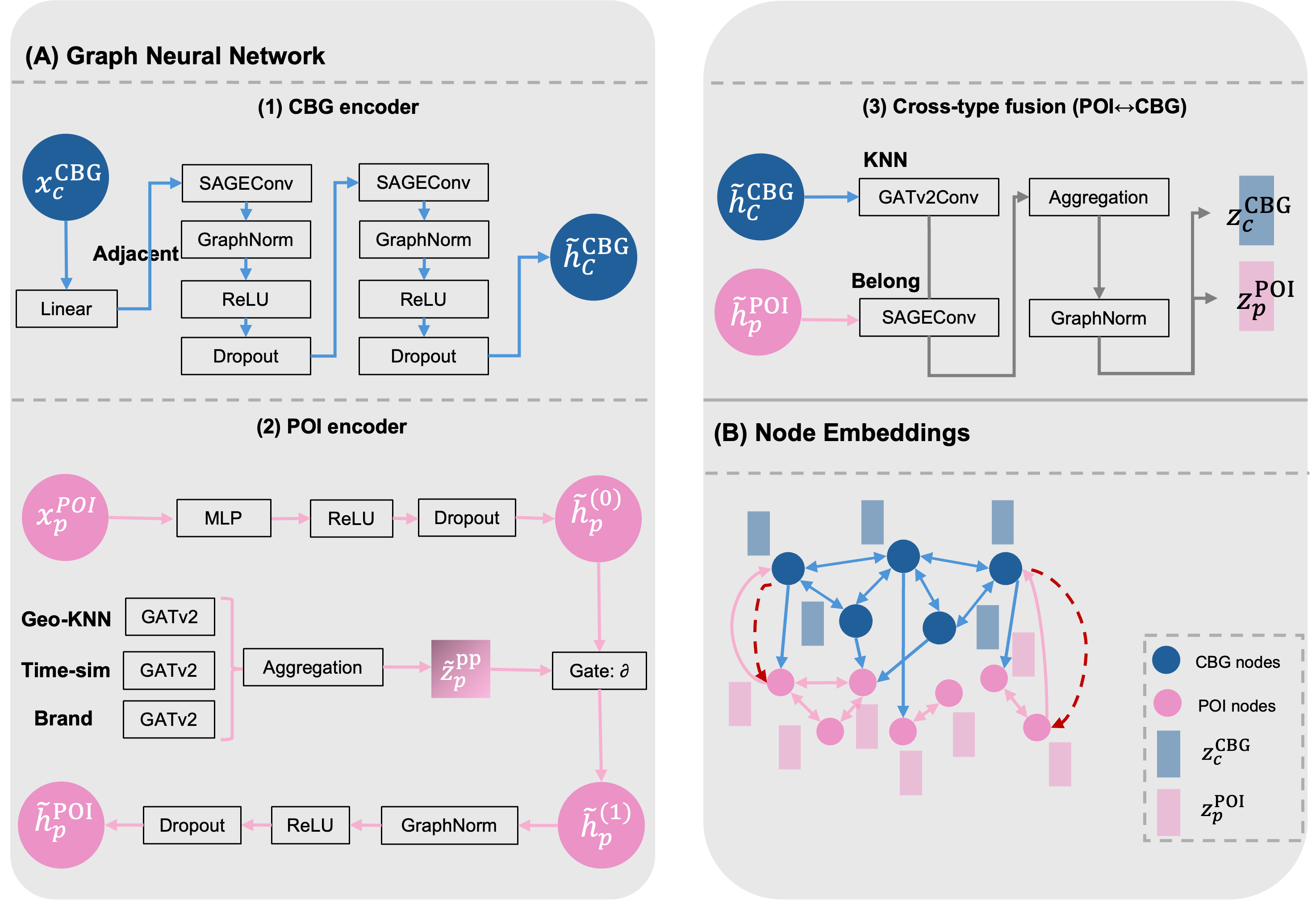}
    \caption{Overview of VisitHGNN}
    \label{Figuregnn_model}
\end{figure}

VisitHGNN is a multi-layer heterogeneous graph neural network designed to predict the likelihood of visits from CBGs to POIs.  The model integrates spatial, functional, and socio-demographic information by jointly encoding both node types and their cross-type relations. The architecture consists of three main components: (1) a CBG encoder that propagates neighborhood context over the CBG graph, incorporating socio-demographic features and adjacency structure; (2) a POI encoder that transforms POI node attributes, including numeric, categorical and text-derived features,  into a compact latent representation, and (3) a cross-type fusion module that exchanges information along POI--CBG edges, allowing contextualized signals from CBGs and POIs to be mutually reinforced. The resulting embeddings are passed to a candidate-constrained pairwise scorer, which evaluates POI--CBG pairs within a restricted candidate set and outputs calibrated visit probabilities. An overview of the full architecture is shown in Figure~\ref{Figuregnn_model}: (\textit{C}) illustrates the heterogeneous message passing; (\textit{D}) highlights the resulting POI and CBG embeddings used by the scoring head.

\subsection{CBG encoder}
\label{subsec:cbg-encoder}
CBG features $x^{\mathrm{CBG}}_{c}\!\in\!\mathbb{R}^{d_{\mathrm{CBG,in}}}$ are linearly projected to $d_{\mathrm{CBG}}$ and refined by two GraphSAGE~\cite{hamilton2018} layers with mean aggregation on the CBG adjacency graph; each layer applies GraphNorm, ReLU, dropout, and a residual connection, as noted by Equations 1-3.
\begin{equation}
h^{(0)}_{c} \;=\; W_{\mathrm{CBG}}\, x^{\mathrm{CBG}}_{c}
\end{equation}

\begin{equation}
\begin{aligned}
m^{(\ell)}_{c} &= \operatorname*{AGG}_{n \in \mathcal{N}_{\mathrm{adj}}(c)}\, W^{(\ell)}_{\mathrm{nbr}}\, h^{(\ell)}_{n},\\
\widehat{h}^{(\ell+1)}_{c} &= \sigma\!\big( W^{(\ell)}_{\mathrm{self}}\, h^{(\ell)}_{c} \;\Vert\; m^{(\ell)}_{c} \big)
\end{aligned}
\end{equation}

\begin{equation}
h^{(\ell+1)}_{c} = Dropout(\operatorname{GraphNorm}(\widehat{h}^{(\ell+1)}_{c})) + h^{(\ell)}_{c}, \quad \ell\in\{0,1\}.
\end{equation}
The encoder output is $\tilde{h}^{\mathrm{CBG}}_{c}\!\coloneqq\! h^{(2)}_{c}\in\mathbb{R}^{d_{\mathrm{CBG}}}$.

\subsection{POI encoder}
\label{subsec:poi-encoder}

We model dependencies between points of interest (POIs) using an encoder that couples a multi-relational POI--POI message-passing block with a global gated residual mechanism \cite{li2015gated,cho2014learning}. Raw POI features $x^{\mathrm{POI}}_{p}\!\in\!\mathbb{R}^{d_{\mathrm{POI,in}}}$ are first projected to a base representation by a two-layer MLP with GraphNorm, ReLU (denoted $\sigma$), and dropout:
\begin{equation}
\tilde{h}^{(0)}_{p}
= Dropout\!\Big(
  \sigma\!\big(
    \operatorname{GraphNorm}\!\big(
      W^{\mathrm{POI}}_{2}\,\sigma\!\big(W^{\mathrm{POI}}_{1}\,x^{\mathrm{POI}}_{p}\big)
    \big)
  \big)
\Big).
\end{equation}
To support heterogeneous interactions $r\!\in\!\mathcal{R}\subseteq\{\texttt{geo\_KNN},\texttt{time\_sim},\texttt{brand}\}$, raw edge attributes $e^{r}_{uv}$ are standardized per relation and then mapped by relation-specific MLPs $\phi_r$ into a common $d_e$-dimensional edge-attribute space:
\begin{equation}
\hat e^{r}_{uv}=\frac{e^{r}_{uv}-\mu_r}{\sigma_r},\quad
a^{r}_{uv}=\phi_r(\hat e^{r}_{uv})\in\mathbb{R}^{d_e}.
\end{equation}

For each relation, a GATv2 layer conditions attention on node and edge information prior to the nonlinearity \cite{brody2021gatv2}, producing attention scores, normalized weights, and messages:
\begin{align}
\ell^{r}_{up} &= \operatorname{LeakyReLU}\!\Big({a_r}^{\!\top}\!\big[\,W^{r}_s\,\tilde h^{(0)}_{u}\;\Vert\;W^{r}_t\,\tilde h^{(0)}_{p}\;\Vert\;U_r\,a^{r}_{up}\big]\Big), \\
\alpha^{r}_{up} &= \operatorname{softmax}_{u\in\mathcal{N}^{r}(p)}\!\ell^{r}_{up}, \\
m^{r}_{p} &= \sum_{u\in\mathcal{N}^{r}(p)}\alpha^{r}_{up}\,M^{r}\tilde h^{(0)}_{u}.
\end{align}

Relation-wise messages are aggregated by $\operatorname{AGG}\!\in\!\{\mathrm{mean},\mathrm{sum},\mathrm{max}\}$ to form
\begin{equation}
z^{\mathrm{pp}}_{p}=\operatorname{AGG}_{r\in\mathcal{R}} m^{r}_{p}.
\end{equation}
A global scalar gate $\alpha=\sigma(\gamma) \in[0,1]$ integrates this signal with the base embedding via a residual update \cite{veit2016residual},
\begin{equation}
\tilde h^{(1)}_{p}=(1-\alpha)\,\tilde h^{(0)}_{p}+\alpha\,z^{\mathrm{pp}}_{p},
\end{equation}
after which GraphNorm, ReLU, and dropout yield the encoder output
\begin{equation}
\tilde h^{\mathrm{POI}}_{p}
=Dropout\!\Big(\sigma\!\big(\operatorname{GraphNorm}(\tilde h^{(1)}_{p})\big)\Big).
\end{equation}

\subsection{Cross-type fusion (POI--CBG)}
\label{subsec:cross-type}

Cross-type fusion operates on the outputs of the POI and CBG encoders, exchanging information along POI--CBG edges. The POI input to the fusion module is the POI-encoder output, denoted $\tilde h^{\mathrm{POI}}_{p}$. The CBG input to the fusion module is
$\tilde h^{\mathrm{CBG}}_{c} \coloneqq \tilde h^{(2)}_{c}$. Heterogeneous graph convolutions are applied over two relation types: (1) \emph{belong} edges, which are processed using a GraphSAGE kernel \cite{hamilton2017inductive}, and (2) \emph{KNN} edges, which are processed using a GATv2 kernel \cite{brody2021gatv2} augmented with a scalar distance feature $d_{p\to c}$. For each POI $p$, messages are aggregated from its CBG neighbors in $\mathcal N_{\text{belong}}(p)$ and $\mathcal N_{\text{KNN}}(p)$. Unless otherwise specified, aggregation is performed via element-wise summation.

\begin{equation}
\widehat z^{\mathrm{POI}}_{p}
=
\sum_{c\in \mathcal N_{\text{belong}}(p)}
\phi_{\text{SAGE}}\!\left(\tilde h^{\mathrm{POI}}_{p},\, \tilde h^{\mathrm{CBG}}_{c}\right)
+
\sum_{c\in \mathcal N_{\text{KNN}}(p)}
\phi_{\text{GATv2}}\!\left(\tilde h^{\mathrm{POI}}_{p},\, \tilde h^{\mathrm{CBG}}_{c},\, d_{p\to c}\right).
\end{equation}

For the CBG side, messages from incident POIs are aggregated similarly:
\begin{equation}
\widehat z^{\mathrm{CBG}}_{c}
=
\sum_{p:\,c\in \mathcal N_{\text{belong}}(p)}
\psi_{\text{SAGE}}\!\left(\tilde h^{\mathrm{CBG}}_{c},\, \tilde h^{\mathrm{POI}}_{p}\right)
+
\sum_{p:\,c\in \mathcal N_{\text{KNN}}(p)}
\psi_{\text{GATv2}}\!\left(\tilde h^{\mathrm{CBG}}_{c},\, \tilde h^{\mathrm{POI}}_{p},\, d_{p\to c}\right).
\end{equation}

The relation kernel for \emph{belong} edges (SAGE) is specified as follows:
\begin{equation}
\phi_{\text{SAGE}}\!\left(\tilde h^{\mathrm{POI}}_{p},\, \tilde h^{\mathrm{CBG}}_{c}\right)
=
W^{\mathrm{bel}}_{\mathrm{self}}\, \tilde h^{\mathrm{POI}}_{p}
\;\Vert\;
W^{\mathrm{bel}}_{\mathrm{nbr}}\, \tilde h^{\mathrm{CBG}}_{c},
\end{equation}
where, a point-wise nonlinearity applied inside the fusion block.  For \emph{KNN} edges (GATv2 with distance),
\begin{equation}
\begin{aligned}
s_{p\to c} &= a^{\top}\,\sigma\!\left(
U\,\tilde h^{\mathrm{POI}}_{p}
\;\Vert\;
V\,\tilde h^{\mathrm{CBG}}_{c}
\;\Vert\;
\omega\, d_{p\to c}
\right).
\end{aligned}
\end{equation}

\begin{equation}
\alpha_{p\to c} = \operatorname{softmax}_{c\in \mathcal N_{\text{KNN}}(p)}\!\left(s_{p\to c}\right).
\end{equation}

\begin{equation}
\phi_{\text{GATv2}}\!\left(\tilde h^{\mathrm{POI}}_{p},\, \tilde h^{\mathrm{CBG}}_{c},\, d_{p\to c}\right)
= \alpha_{p\to c}\, W^{\mathrm{KNN}}\, \tilde h^{\mathrm{CBG}}_{c}.
\end{equation}
we apply GraphNorm to the fused representations to yield the node embeddings for the downstream task.
\begin{equation}
z^{\mathrm{POI}}_{p} = \operatorname{GraphNorm}\!\left(\widehat z^{\mathrm{POI}}_{p}\right),
\qquad
z^{\mathrm{CBG}}_{c} = \operatorname{GraphNorm}\!\left(\widehat z^{\mathrm{CBG}}_{c}\right).
\end{equation}

When a node receives no cross-type messages in the current minibatch (e.g., $\mathcal N_{\text{belong}}(p)\cup\mathcal N_{\text{KNN}}(p)=\emptyset$), its encoder output is passed through a linear projection to the shared hidden dimension $d_{\mathrm{hid}}$, ensuring dimensional consistency across node types, even in the absence of active cross-type neighbors.

\subsection{Node embeddings}
\label{subsec:node-emb}
Following the encoders and cross-type fusion, each POI $p$ and CBG $c$ is represented by a $d_{\mathrm{hid}}$-dimensional vector,
\[
z^{\mathrm{POI}}_{p}\in\mathbb{R}^{d_{\mathrm{hid}}},\qquad
z^{\mathrm{CBG}}_{c}\in\mathbb{R}^{d_{\mathrm{hid}}}.
\]
The inputs to the fusion module are the encoder outputs: for POIs, $\tilde h^{\mathrm{POI}}_{p}$; for CBGs, $\tilde h^{\mathrm{CBG}}_{c}\coloneqq h^{(2)}_{c}$. When a node receives no cross-type messages in the current minibatch (e.g., $\mathcal N_{\text{belong}}(p)\cup\mathcal N_{\text{KNN}}(p)=\emptyset$), its encoder output is passed through a fallback linear projection into $d_{\mathrm{hid}}$, ensuring consistent embedding dimensionality across node types. These embeddings serve as inputs to the prediction head in Section~\ref{sec:pred}. For each POI $p$ and its candidate origin set $N_{K}(p)$, we construct pairwise features via concatenation,
\[
\mathbf{F}_{p}
=
\big[\, z^{\mathrm{CBG}}_{c_1}\!\Vert\, z^{\mathrm{POI}}_{p};\;
      \ldots;\;
      z^{\mathrm{CBG}}_{c_K}\!\Vert\, z^{\mathrm{POI}}_{p}\,\big]
\in \mathbb{R}^{K\times 2d_{\mathrm{hid}}},
\]
which are then passed to the MLP scorer (see Section~\ref{subsec:head}).

\section{Prediction and loss function}
\label{sec:pred}

\begin{figure}[H]
    \centering
    \includegraphics[width=1\linewidth]{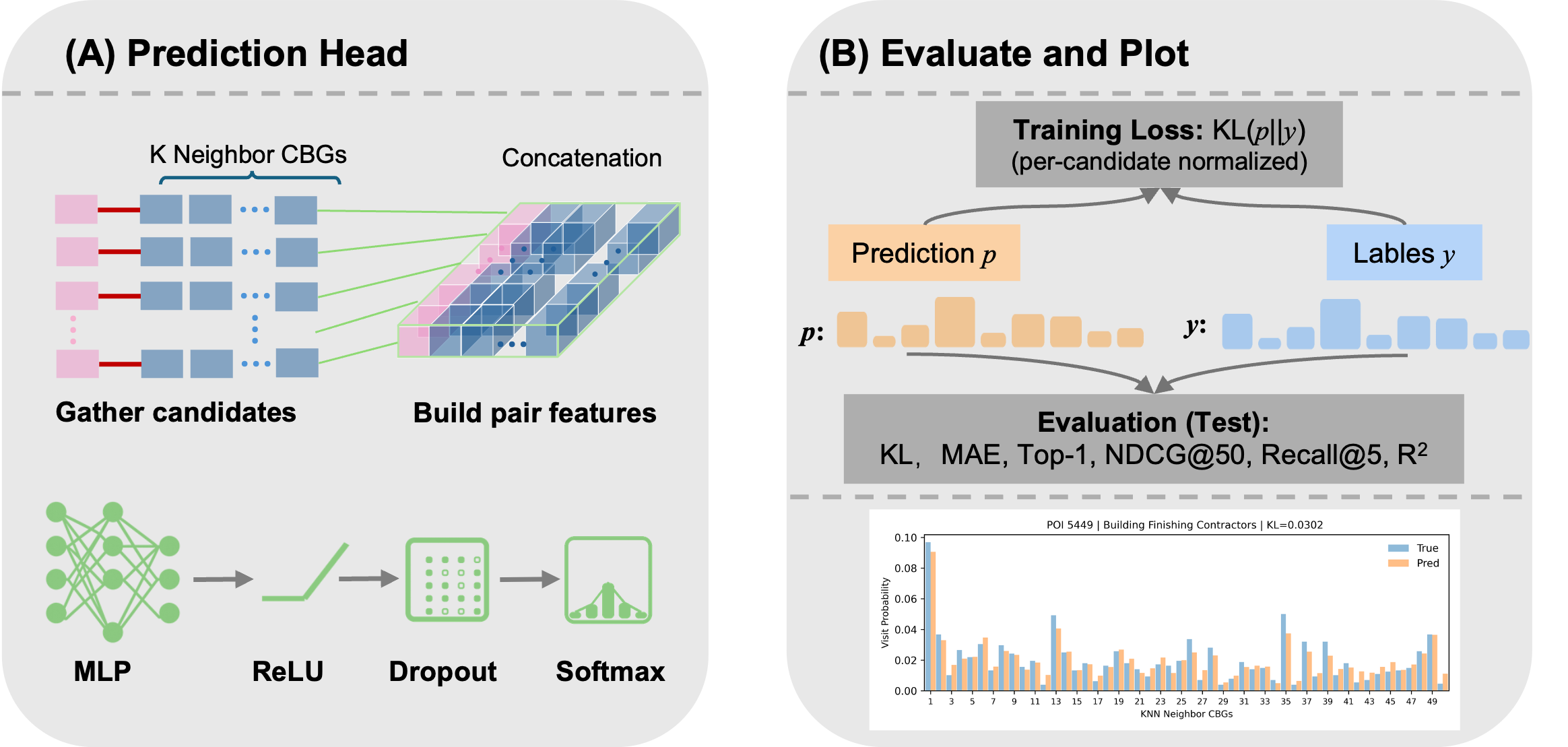}
    \caption{Prediction and evaluation pipeline in VisitHGNN}
    \label{Figurepred_head}
\end{figure}

This section details the prediction layer of VisitHGNN, which estimates, for each POI $p$, the origin distribution of visits over a spatially constrained candidate set of CBGs. Given the contextual embeddings produced by the encoders and cross-type fusion (Section~\ref{sec:method}), we form concatenated POI--CBG pair features and score them with a shared multilayer perceptron (MLP). To accommodate variable candidate-set sizes, probabilities are obtained with a masked softmax. Training minimizes a masked Kullback--Leibler (KL) divergence between predicted and observed visit distributions. Evaluation uses both ranking- and calibration-oriented metrics to assess accuracy and reliability. Figure~\ref{Figurepred_head}(B) includes a compact plot inset that, for a representative POI $p$, juxtaposes the predicted distribution $p_{p,\cdot}$ with the observed distribution $y_{p,\cdot}$ over the $K$ candidates $N_{K}(p)$.

\subsection{Candidate CBG set for each POI}
\label{subsec:candidates}
The neighborhood size $K$ controls the spatial extent of candidate CBGs considered for each POI. For a given POI $p$, the candidate set $N_{K}(p)$ comprises the $K$ nearest CBGs under Euclidean distance in a projected UTM coordinate system. This design bounds the prediction space to plausible origins while remaining computationally tractable. Rather than using a fixed global value, $K$ is tuned within a graph-specific range that scales with the number of CBGs in the region ($|\mathcal{C}|$), balancing locality and coverage. In practice, $K$ is selected via Optuna using validation masked KL divergence as the objective. Once $N_{K}(p)$ is fixed, it is held constant for training and evaluation to ensure reproducibility and to isolate the effects of evolving node features and relations.

\subsection{Pairwise scoring head}
\label{subsec:head}
Let $N_{K}(p)=\{c_{1},\dots,c_{K}\}$. Using the embeddings from Section~\ref{sec:method},
$z^{\mathrm{POI}}_{p}, z^{\mathrm{CBG}}_{c}\in\mathbb{R}^{d_{\mathrm{hid}}}$, we build POI--CBG pair features by concatenation and score each pair with a shared MLP $f_{\theta}$:
\begin{equation}
s_{p,k} \;=\; f_{\theta}\!\big(z^{\mathrm{CBG}}_{c_{k}} \,\Vert\, z^{\mathrm{POI}}_{p}\big), 
\qquad k=1,\ldots,K.
\end{equation}
Stacking $\{s_{p,k}\}_{k=1}^{K}$ yields logits $s_{p}\in\mathbb{R}^{K}$. In implementation, $f_{\theta}$ is a feed-forward network with ReLU activations and dropout, and the parameters $\theta$ are shared across all POIs.

\subsection{Masked softmax and loss}
\label{subsec:loss}
Let $m_{p,k}\in\{0,1\}$ indicate whether the $k$-th position corresponds to a valid candidate in $N_{K}(p)$, and let $y_{p,k}$ denote the observed visit distribution for POI $p$, normalized over $N_{K}(p)$. We compute probabilities via a masked softmax
\begin{equation}
p_{p,k}
\;=\;
\frac{\exp(s_{p,k})\,m_{p,k}}{\sum_{j=1}^{K}\exp(s_{p,j})\,m_{p,j}},
\end{equation}
and minimize the masked KL divergence over training POIs $\mathcal{P}_{\mathrm{tr}}$:
\begin{equation}
\mathcal{L}
\;=\;
\frac{1}{|\mathcal{P}_{\mathrm{tr}}|}
\sum_{p\in\mathcal{P}_{\mathrm{tr}}}
\sum_{k=1}^{K}
y_{p,k}\,\log\!\frac{y_{p,k}}{p_{p,k}}\,m_{p,k}.
\end{equation}
Model performance is reported on held-out validation and test POIs in a transductive setting, where supervision edges are withheld for non-training POIs. Metrics are computed only over valid candidates and include ranking measures (Top-1 accuracy, recall at various cutoffs) and calibration/regression measures such as mean absolute error (MAE) and the coefficient of determination ($R^{2}$).

\section{Experiments}\label{sec:experiments}

\subsection{Experimental Setup}\label{subsec:setup}
To demonstrate the efficacy of our approach, we evaluate it on the Dewey Weekly Patterns dataset \cite{Dewey} for Fulton County, Georgia, using the first week of January 2019. From 58,757 venues, we sample 8,000 POIs with sufficient visit activity to support the supervised task of predicting visitor-origin distributions. All models are trained and evaluated using a deterministic, transductive split of POI nodes (70\% train, 15\% validation, 15\% test). Model selection is based on the lowest validation KL divergence, and test results are reported.
\\
\subsubsection{Graph construction and candidates}
We construct a heterogeneous graph with CBG–CBG adjacency \((\texttt{cbg, adjacent, cbg})\), cross-type POI--CBG relations (\texttt{belong} and distance-based \texttt{KNN}, plus reverse forms), and multi-relational POI–POI edges capturing three complementary channels: \texttt{geo\_KNN},\texttt{time\_sim},\texttt{brand}, represented by a unified edge-attribute vector. At the graph building stage, we precompute distances from each POI to \emph{all} CBGs in Fulton County, yielding a full candidate list per POI. During training and evaluation, we truncate each list to the top-$K$ nearest CBGs, with $K$ tuned by Optuna over the set $\{10,12,14,\dots,50\}$.

\subsubsection{Nodes and Features}
POI and CBG nodes share a locked, order-stable schema to guarantee strict comparability across runs. POI features include engineered attributes capturing geometric properties and activity intensity, aggregated opening-hours statistics, and categorical encodings. Two optional augmentations are evaluated: (1) text embeddings obtained from a BERT-base model using place names and descriptors, and (2) normalized dwell-time histograms. CBG features are assembled in an analogous manner using a fixed numerical schema. Missing attributes are zero-imputed and all node features are row-wise normalized.

\subsubsection{Model training results}
\begin{figure}[H]
    \centering
    \includegraphics[width=0.95\linewidth]{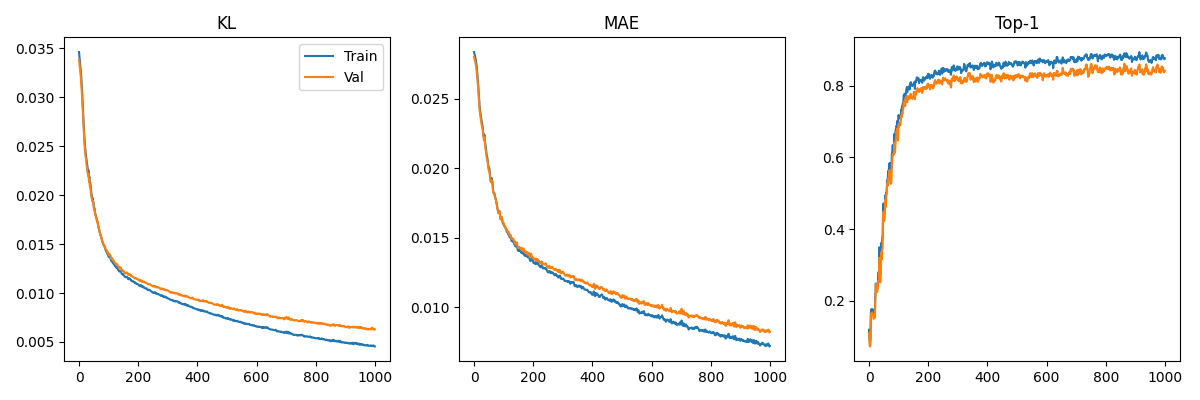}
    \caption{Training dynamics of VisitHGNN.}
    \label{Figuretraining-curves}
\end{figure}

Training was conducted in a transductive setting on the entire heterogeneous graph. The loss is computed as a masked, row-wise KL divergence over shared KNN candidate sets; for training stability, this loss is additionally normalized by the number of candidates (\(K\)), which scales the metric differently than the normalized KL used in final evaluations. Model selection was based on the minimum validation loss. As shown in Figure~\ref{Figuretraining-curves}, the training curves exhibit rapid early improvement followed by a smooth plateau: the validation KL divergence decreases from approximately \(3.5 \times 10^{-2}\) to \(5 \times 10^{-3}\), while the MAE declines from \(\sim 2.8 \times 10^{-2}\) to \(\sim 6 \times 10^{-3}\). In parallel, Top-1 accuracy rose sharply from near-chance levels to \(\approx 0.88\)–\(0.90\). The training and validation trajectories remain closely aligned with a modest, stable gap and no late-epoch degradation. This indicates stable convergence without signs of overfitting and establishes a solid foundation for subsequent comparative evaluation.

\subsubsection{Evaluation Metrics}
For each POI, the model estimates a probability distribution over a fixed candidate set of CBGs derived from distance and truncated to size \(K\). The target distributions are defined as row-normalized visit counts. Under the transductive split (Section~\ref{subsec:split}), each epoch performs message passing and forward computation on the full graph; the loss is computed only on training POIs, and validation and test labels are excluded from the loss and used solely for evaluation. Model checkpoints are selected based on validation KL divergence \(D_{\mathrm{KL}}\), and test metrics are reported for the checkpoint yielding the lowest validation \(D_{\mathrm{KL}}\).

Table~\ref{tab:metrics} lists the metrics used to evaluate model performance. KL divergence \(D_{\mathrm{KL}}(t_i \,\|\, p_i)\) serves as the primary metric, as it compares the full predicted distribution \(p_i\) with the observed distribution \(t_i\) and penalizes probability mass assigned to incorrect CBGs, particularly those never visited. Mean absolute error (MAE) captures the average absolute difference between the predicted and observed probabilities across each POI's candidate CBG set. \(\mathrm{Top}\text{-}1 \) accuracy assesses whether the highest-probability CBG predicted matches the most likely CBG in the ground truth. Normalized Discounted Cumulative Gain at \textit{k} (\(\mathrm{NDCG}@k\)) evaluates the model's ability to rank relevant CBGs correctly, considering their probability and position in the ranked list up to the top \textit{k}. $\mathrm{rank}_i(j)$ is the rank of candidate $j$ under predicted scores for POI $i$. \(\mathrm{Recall}@k\) quantifies the proportion of true  probability mass contained within the top-k predictions. Finally, \(R^2\) reflects overall goodness-of-fit across all valid POI--CBG pairs after row normalization. As indicated in the table, lower values are preferable for \(D_{\mathrm{KL}}\) and MAE, whereas higher values denote better performance for all other metrics.

\begin{table}[H]
\centering
\small
\caption{Evaluation metrics}
\label{tab:metrics}
\renewcommand{\arraystretch}{2.2}
\setlength{\extrarowheight}{2pt} 
\begin{tabular}{@{}l l p{8cm} c c@{}}
\toprule
Metric & Notation & Definition & Range & Better \\
\midrule
KL divergence & $D_{\mathrm{KL}}(t_i\|p_i)$
& $\displaystyle \sum_{j \in \mathcal{C}_i} t_{ij}\,\log\!\frac{t_{ij}+\epsilon}{p_{ij}+\epsilon}$, averaged over POIs.
& $[0,\infty)$ & $\downarrow$ \\
Mean absolute error & $\mathrm{MAE}(i)$
& $\displaystyle \frac{1}{|\mathcal{C}_i|}\sum_{j \in \mathcal{C}_i} \big|p_{ij}-t_{ij}\big|$, averaged over POIs.
& $[0,1]$ & $\downarrow$ \\
Top-1 accuracy & $\mathrm{Top\text{-}1}$ 
& $\displaystyle \frac{1}{|\mathcal{I}|}\sum_{i\in\mathcal{I}}
\mathbf{1}\!\Big[
\arg\max_{j\in\mathcal{C}_i} p_{ij}\ \in\ \operatorname*{Arg\,max}_{j\in\mathcal{C}_i} t_{ij}
\Big]$
& $[0,1]$ & $\uparrow$ \\

NDCG@$k$ & $\mathrm{NDCG@}k$
& $\displaystyle \frac{1}{Z_i}\sum_{j \in \mathrm{top}\,k(p_i)} \frac{t_{ij}}{\log_2\!\big(1+\mathrm{rank}_i(j)\big)}$, where $Z_i$ is the ideal DCG computed from $t_i$.
& $[0,1]$ & $\uparrow$ \\
Recall@$k$ & $\mathrm{Recall@}k$
& $\displaystyle \frac{\sum_{j \in \mathrm{top}\,k(p_i)} t_{ij}}{\sum_{j \in \mathcal{C}_i} t_{ij}}$, averaged over POIs with $\sum_{j \in \mathcal{C}_i} t_{ij}>0$.
& $[0,1]$ & $\uparrow$ \\
Coefficient of determination& $R^2$
& $\displaystyle 1-\frac{\sum_{\!(i,j)\ \text{valid}} \big(t_{ij}-p_{ij}\big)^2}{\sum_{\!(i,j)\ \text{valid}} \big(t_{ij}-\bar{t}\big)^2}$, computed over all valid POI--CBG pairs after row-normalization ($\bar{t}$ is the global mean of $t_{ij}$).
& $(-\infty,1]$ & $\uparrow$ \\
\bottomrule
\end{tabular}
\vspace{0.25em}
\end{table}

\subsection{Ablation Studies}\label{subsec:ablations}
We conducted systematic ablation experiments to assess the contribution of individual structural and architectural components of the model, focusing on the four design components summarized in Table~\ref{tab:primary_components}.

\begin{table}[H]
\centering
\caption{Primary design components evaluated (panels correspond to Figure~\ref{Figureablation}).}
\begin{tabular}{@{}lll@{}}
\toprule
Panel & Suite & Description \\
\midrule
(a) & POI--POI relations & Three types of edges: geo\_KNN, time\_sim, brand \\
(b) & CBG--CBG adjacency & Edges between neighboring CBGs \\
(c) & Cross POI--CBG edges & Belong \& attribute-aware KNN variants \\
(d) & GraphNorm & Graph normalization on/off \\
\bottomrule
\end{tabular}
\label{tab:primary_components}
\end{table}

\begin{figure}[H]
    \centering
    \includegraphics[width=1\linewidth]{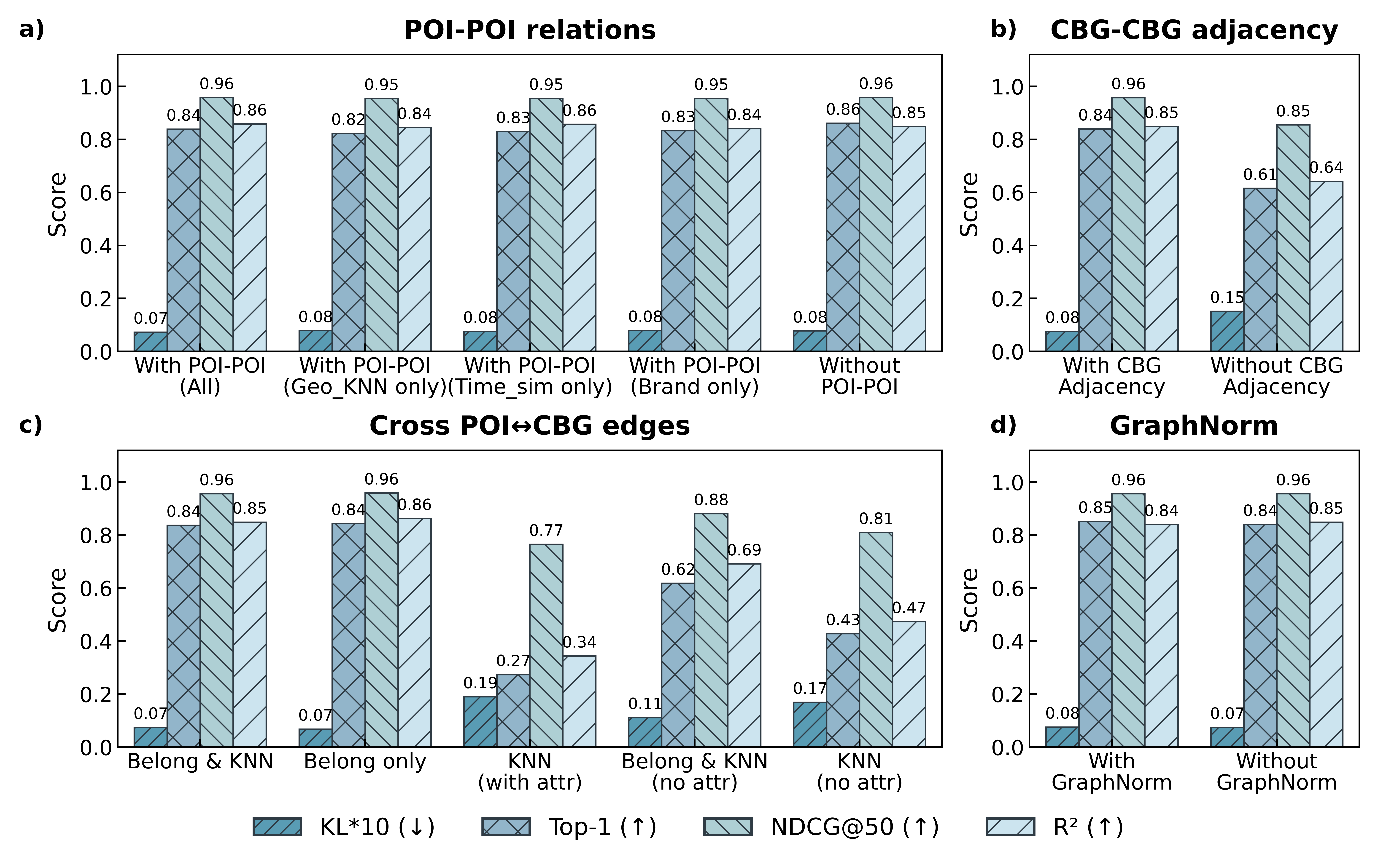}
    \caption{Ablation of graph structure and normalization.}
    \label{Figureablation}
\end{figure}

Removing all POI--POI relations substantially degraded performance relative to the full model, which includes geo\_KNN, time\_sim, and brand links. Retaining any single link partially recovered performance, with geo\_KNN and time\_sim delivering the largest gains, whereas brand alone was insufficient. This indicates that POI structural similarity provides complementary rather than sufficient signal. Eliminating CBG--CBG adjacency further reduced Top-1 accuracy and NDCG@50 while increasing $D_{\mathrm{KL}}$, consistent with spatial smoothing across neighboring CBGs alleviating supervision sparsity; in Figure~\ref{Figureablation}, $D_{\mathrm{KL}}$ is plotted as $\times 10$ for readability. For Cross POI--CBG edges, combining Belong and KNN yielded the strongest results; restricting the graph to a single edge---either Belong only or KNN with attributes---produced measurable declines. Mechanism-focused ablations confirmed that removing edge attributes---Belong and KNN without attributes, and KNN without attributes---degraded performance, underscoring the value of attribute-aware edges. Disabling GraphNorm degraded all metrics, highlighting its role in optimization stability. Overall, bidirectional, attribute-aware Cross POI--CBG edges are most critical, with POI--POI relations and CBG--CBG adjacency providing complementary benefits.

\subsection{Overall Results}\label{subsec:overall}

On the test dataset with \(K=50\) nearest candidate CBGs per POI, we evaluate predicted distribution after masking invalid entries and setting \(\epsilon=10^{-9}\) in computing KL divergence (see Table~\ref{tab:metrics}). For each POI \(i\), \(t_{ij}\) and \(p_{ij}\) denote the observed and predicted probabilities for candidate \(j \in \mathcal{C}_i\). The model shows low divergence from observed visit distributions (mean KL \(= 0.287\)) and small absolute error (MAE \(= 0.008\)). Ranking performance is high: Top-1 accuracy reaches \(0.853\) and NDCG@50 is \(0.966\), indicating close alignment between predicted and empirical orderings and accurate identification of the primary destination for most POIs. The top five predictions recover a substantial portion of the true probability mass (Recall@5 \(= 0.611\)), suggesting residual dispersion beyond the highest-ranked entries. Collectively, these results indicate that the model yields well-fit probability estimates while preserving the observed ranking structure across candidates.

\begin{figure}[H]
    \centering
    \includegraphics[width=0.7\linewidth]{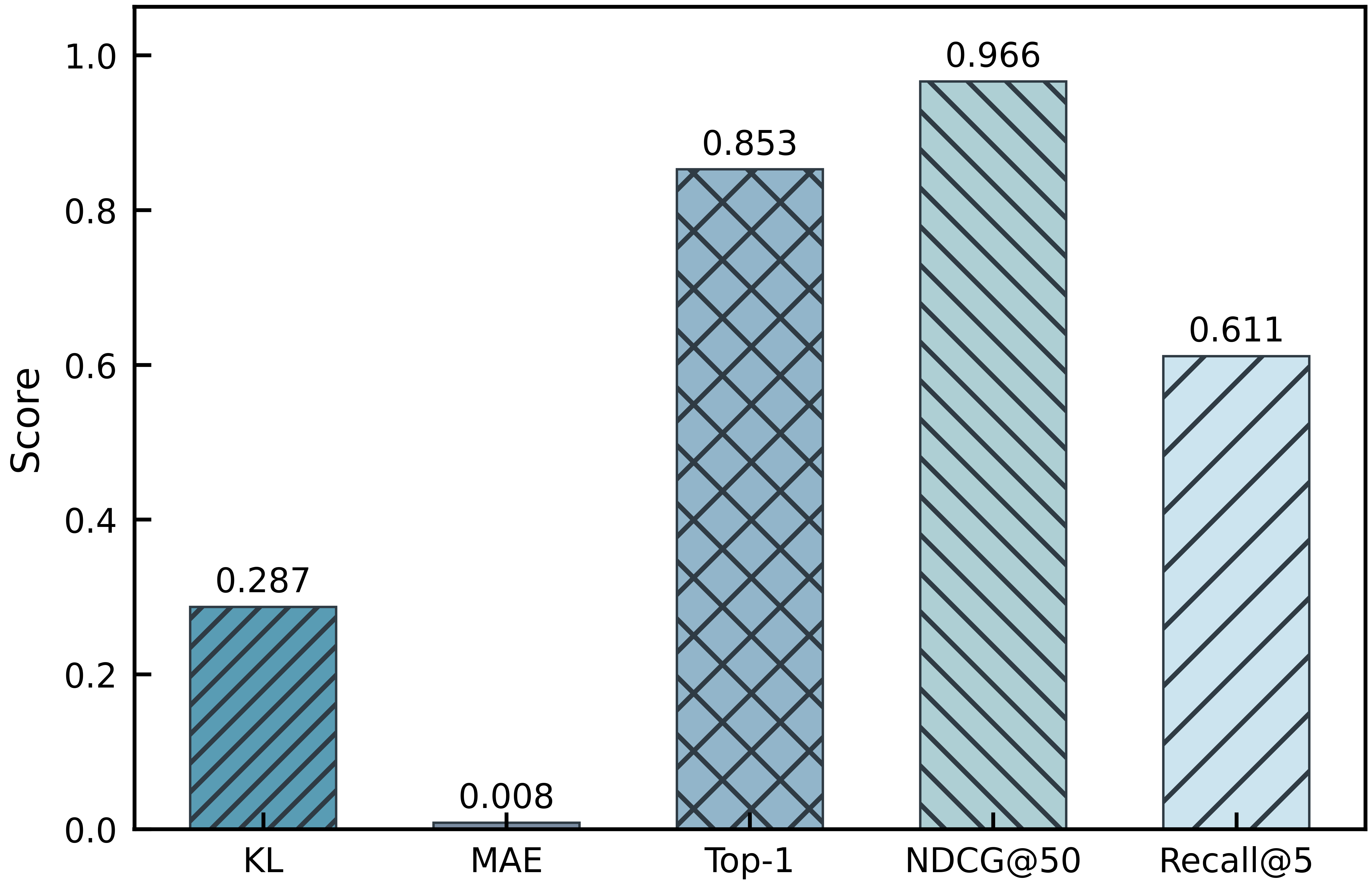}
    \caption{Overall evaluation metrics on the \textit{test} split }
    \label{Figureoverall-metrics-test}
\end{figure}

\begin{figure}[H]
    \centering
    \includegraphics[width=0.7\linewidth]{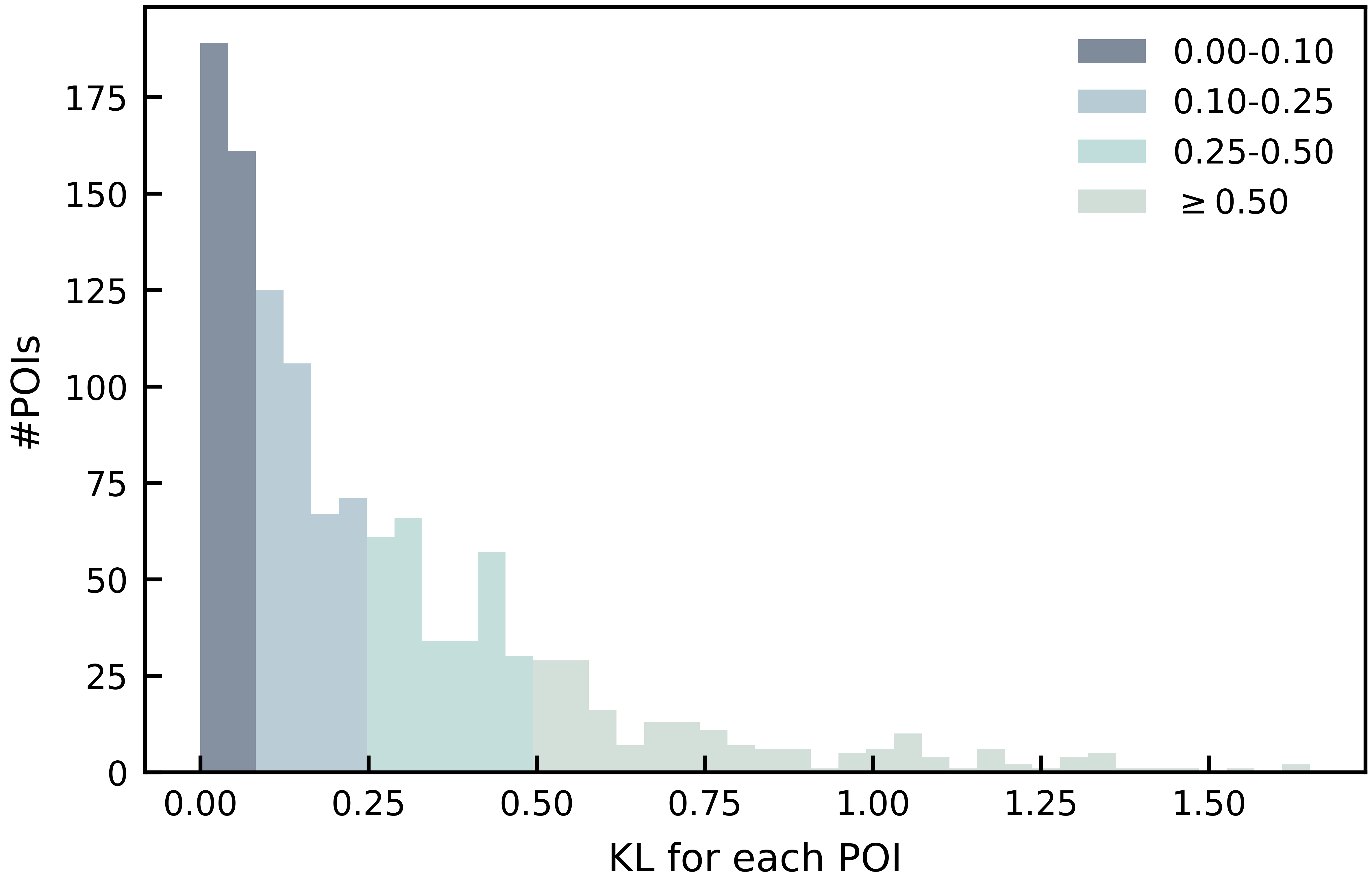}
    \caption{KL divergence across POIs on the test split, $K{=}50$.}
    \label{Figurekl-hist-test}
\end{figure}

Error is not distributed uniformly across POIs. The per-POI KL histogram is sharply peaked near zero with a long right tail, indicating that the majority of POIs are predicted with very small divergence, while a smaller subset exhibits larger mismatches (Figure~\ref{Figurekl-hist-test}). This distribution is consistent with the low average KL and suggests that a few hard cases accounted for the most of the aggregate error under the fixed candidate set evaluation protocol (Section~\ref{subsec:setup}).

\begin{figure}[H]
    \centering
    \includegraphics[width=0.55\linewidth]{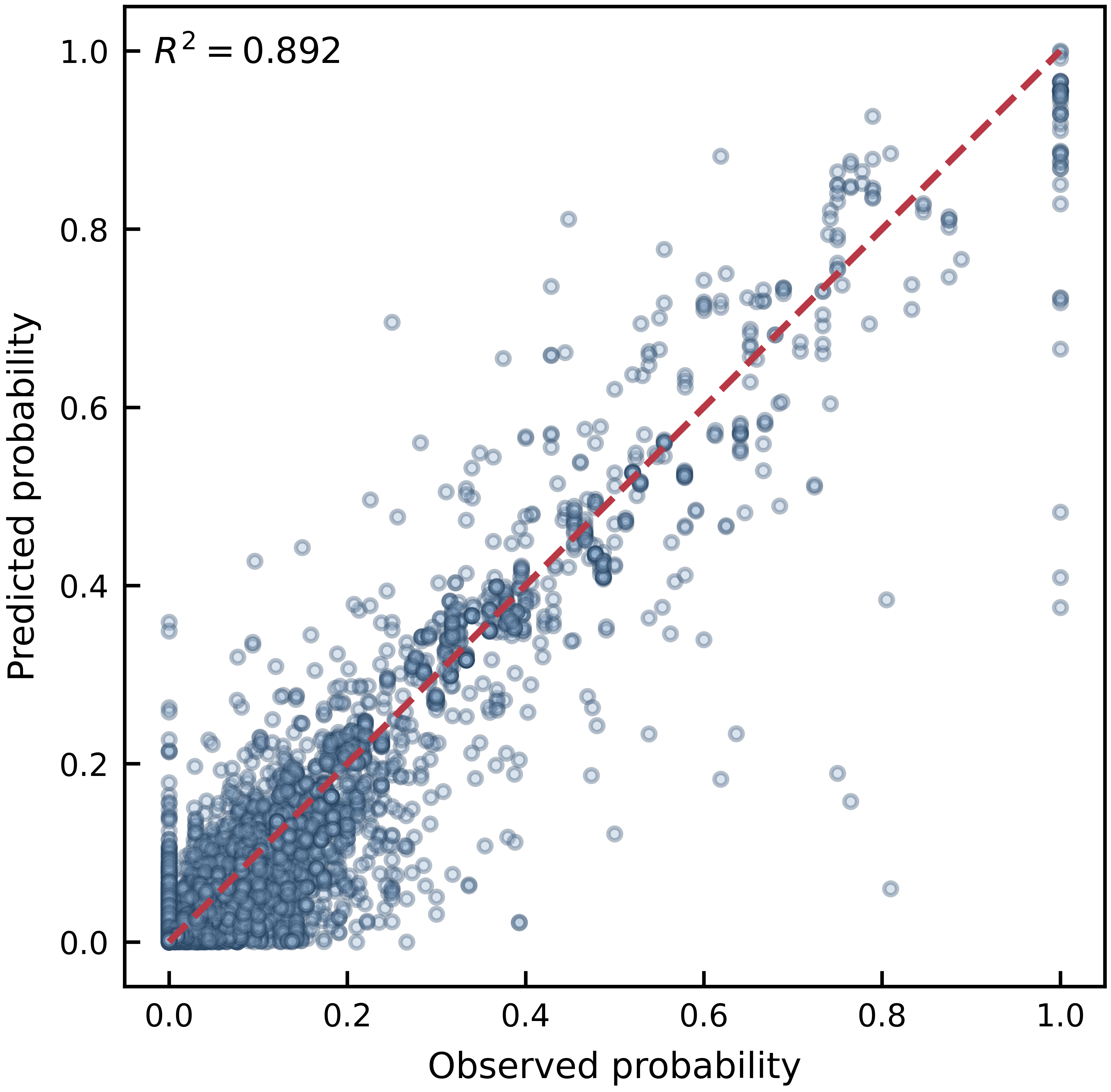}
    \caption{Predicted versus observed visit probabilities for POI--CBG candidate pairs on the test split ($K{=}50$).}
    \label{Figurescatter-test}
\end{figure}

The distribution fitting is further examined by  scatter plot of predicted versus observed probabilities across all valid POI--CBG pairs, with $R^{2}{=}\mathbf{0.892}$ (Figure~\ref{Figurescatter-test}). This high coefficient of determination shows a high level of agreement between observed and predicted probabilities.

\subsection{Model comparison}\label{subsec:comparison}
Using a shared candidate set and ground truth (Table~\ref{tab:overall-comparison}), we benchmarked three approaches: VisitHGNN, a pairwise MLP scorer, and a geographic nearest-neighbor heuristic (KNN-Geo), under an identical in-candidate evaluation protocol. VisitHGNN achieved the best predictions, yielding the lowest divergence from the observed distributions (KL = 0.287) and error (MAE = 0.0082), the highest pointwise accuracy (Top-1 = 0.853), and strong goodness of fit ($R^2$ = 0.892). It also captured a substantial portion of the true probability mass among the highest-ranked candidates (Recall@5/10/15 = 0.611/0.771/0.857). The pairwise MLP was competitive but consistently weaker (KL = 0.643; MAE = 0.0156; Top-1 = 0.744; $R^{2}$ = 0.589; Recall@5/10/15 = 0.570/0.701/0.780). The KNN-Geo baseline, which relies solely on spatial proximity, performed substantially worse (KL = 1.016; MAE = 0.0223; Top-1 = 0.518; $R^{2}$ = 0.099; Recall@5/10/15 = 0.515/0.647/0.716), indicating that geometric distance alone is insufficient to explain visit behavior. Collectively, these results show that combining learned representations of POIs and contextual CBG features within a heterogeneous graph neural network yields significant gains in modeling POI visit patterns compared to both simpler learning-based and heuristic baselines.

\begin{table}[H]
  \centering
  \small
  \caption{Comparison of model performance.}
  \label{tab:overall-comparison}
  \setlength{\tabcolsep}{6pt}
  \renewcommand{\arraystretch}{1.1}
  \begin{tabular}{@{}lccccccc@{}}
    \toprule
    Method & KL & MAE & Top-1 & $R^{2}$ & Rec@5 & Rec@10 & Rec@15 \\
    \midrule
    MLP            & 0.643 & 0.0156 & 0.744 & 0.589 & 0.570 & 0.701 & 0.780 \\
    KNN(Geo)       & 1.016 & 0.0223 & 0.518 & 0.099 & 0.515 & 0.647 & 0.716 \\
    VisitHGNN & 0.287 & 0.0082 & 0.853 & 0.892 & 0.611 & 0.771 & 0.857 \\
    \bottomrule
  \end{tabular}
\end{table}

\section{Discussion and conclusion}
\label{sec:discussion}

This study introduced VisitHGNN, a heterogeneous graph learning framework for modeling POI visit patterns by their origin CBGs. By jointly capturing three types of relationships: POI–POI, CBG–CBG, and POI–CBG, in a unified graph setting, VisitHGNN achieves superior predictive performance. Evaluation on a representative county-scale dataset shows consistent gains over strong baselines, reducing divergence and error while improving Top-1 ranking accuracy. Ablation studies further highlight the importance of its design choices: a multi-channel graph structure, cross-type fusion enriched with distance information, and principled normalization. Together, these components yield origin predictions that are both accurate and operationally meaningful.

The findings reinforce prior evidence that structured deep learning on urban graphs uncovers dependencies that traditional geometric or tabular models struggle to capture. By representing geospatial proximity, temporal co-activity, and brand co-visitation as distinct relational channels, VisitHGNN operationalizes well-established behavioral mechanisms (movement frictions, behavioral synchrony, and functional affinity). Through candidate-restricted inference and a calibrated scoring head, these mechanisms are translated into probability distributions that substantially outperform nearest-neighbor heuristics and point-estimate regressors commonly used in practice. This advance not only improves methodological rigor but also enhances interpretability and transferability.

From a practical perspective, calibrated probability distributions, rather than single-point forecasts, support a broader range of decision-making tasks. They directly enable risk-aware siting, coverage planning, and staffing, while also facilitating transfer across regions and service providers. The modular graph design further accommodates new relational layers (e.g., transit connectivity, roadway impedance), edge attributes, and metadata without requiring architectural redesign, making the framework adaptable to diverse deployment contexts. Nonetheless, several limitations remain. Current evaluation is transductive, candidate-restricted, and limited to a single county–week, underscoring the need for inductive and cross-domain validation across held-out POIs, weeks, and cities. Reliance on panel and administrative sources may also introduce sampling bias and category noise, motivating future work on bias auditing, auxiliary data triangulation, and causal-sensitivity analyses. At national scale, memory and compute demands for message passing and scoring will necessitate scalable solutions such as neighborhood sampling, graph partitioning, sparse or quantized operators, and distributed inference.

In summary, VisitHGNN provides a scalable and practical approach to generating high-resolution, decision-ready origin maps. By integrating heterogeneous relations, constraining inference to plausible neighborhoods, and producing calibrated probability estimates, it advances both methodological accuracy and operational utility for urban and mobility analytics.

\bibliography{references} 

\begin{thebibliography}{35}
\providecommand{\natexlab}[1]{#1}
\providecommand{\url}[1]{\texttt{#1}}
\expandafter\ifx\csname urlstyle\endcsname\relax
  \providecommand{\doi}[1]{doi: #1}\else
  \providecommand{\doi}{doi: \begingroup \urlstyle{rm}\Url}\fi

\bibitem[González et~al.(2008)González, Hidalgo, and Barabási]{Gonzalez2008Understanding}
M.~C. González, C.~A. Hidalgo, and A.-L. Barabási.
\newblock Understanding individual human mobility patterns.
\newblock \emph{Nature}, 453\penalty0 (7196):\penalty0 779--782, 2008.

\bibitem[Song et~al.(2010)]{Song2010Limits}
C.~Song et~al.
\newblock Limits of predictability in human mobility.
\newblock \emph{Science}, 327\penalty0 (5968):\penalty0 1018--1021, 2010.

\bibitem[Rong et~al.(2025)Rong, Ding, and Li]{Rong2025SurveyOD}
C.~Rong, J.~Ding, and Y.~Li.
\newblock An interdisciplinary survey on origin-destination flows modeling: Theory and techniques.
\newblock \emph{ACM Computing Surveys}, 57\penalty0 (1):\penalty0 1--49, 2025.

\bibitem[Chang et~al.(2021)]{Chang2021MobilityCOVID}
S.~Chang et~al.
\newblock Mobility network models of covid-19 explain inequities and inform reopening.
\newblock \emph{Nature}, 589\penalty0 (7840):\penalty0 82--87, 2021.

\bibitem[Jung et~al.(2008)Jung, Wang, and Stanley]{Jung2008GravityKoreanHighway}
W.-S. Jung, F.~Wang, and H.~E. Stanley.
\newblock Gravity model in the korean highway.
\newblock \emph{Europhysics Letters}, 81\penalty0 (4):\penalty0 48005, 2008.

\bibitem[Stouffer(1940)]{Intervening}
Samuel~A. Stouffer.
\newblock Intervening opportunities: A theory relating mobility and distance.
\newblock \emph{American Sociological Review}, 5\penalty0 (6):\penalty0 845--867, 1940.
\newblock ISSN 00031224.
\newblock URL \url{http://www.jstor.org/stable/2084520}.

\bibitem[Simini et~al.(2012)]{Simini2012UniversalModel}
F.~Simini et~al.
\newblock A universal model for mobility and migration patterns.
\newblock \emph{Nature}, 484\penalty0 (7392):\penalty0 96--100, 2012.

\bibitem[Ren et~al.(2014)]{Ren2014RadiationTemporal}
Y.~Ren et~al.
\newblock Predicting commuter flows in spatial networks using a radiation model based on temporal ranges.
\newblock \emph{Nature Communications}, 5:\penalty0 5347, 2014.

\bibitem[Lenormand et~al.(2015)]{Lenormand2015SocioDemoMobility}
M.~Lenormand et~al.
\newblock Influence of sociodemographic characteristics on human mobility.
\newblock \emph{Scientific Reports}, 5:\penalty0 10075, 2015.

\bibitem[Rodríguez-Rueda et~al.(2021)]{RodriguezRueda2021ODMatrixML}
P.~J. Rodríguez-Rueda et~al.
\newblock Origin--destination matrix estimation and prediction from socioeconomic variables using automatic feature selection procedure-based machine learning model.
\newblock \emph{Journal of Urban Planning and Development}, 147\penalty0 (4), 2021.

\bibitem[Simini et~al.(2021)]{Simini2021DeepGravity}
F.~Simini et~al.
\newblock A deep gravity model for mobility flows generation.
\newblock \emph{Nature Communications}, 12\penalty0 (1), 2021.

\bibitem[Robinson and Dilkina(2018)]{Robinson}
Caleb Robinson and Bistra Dilkina.
\newblock A machine learning approach to modeling human migration.
\newblock In \emph{Proceedings of the 1st ACM SIGCAS Conference on Computing and Sustainable Societies}, COMPASS '18, New York, NY, USA, 2018. Association for Computing Machinery.
\newblock ISBN 9781450358163.
\newblock \doi{10.1145/3209811.3209868}.
\newblock URL \url{https://doi.org/10.1145/3209811.3209868}.

\bibitem[Cai et~al.(2022)Cai, Pang, and Sekimoto]{Cai2022SpatialAttentionOD}
M.~Cai, Y.~Pang, and Y.~Sekimoto.
\newblock Spatial attention based grid representation learning for predicting origin--destination flow.
\newblock In \emph{2022 IEEE International Conference on Big Data (Big Data)}. IEEE, 2022.

\bibitem[Pourebrahim et~al.(2019)]{Pourebrahim2019TripTwitter}
N.~Pourebrahim et~al.
\newblock Trip distribution modeling with twitter data.
\newblock \emph{Computers, Environment and Urban Systems}, 77:\penalty0 101354, 2019.

\bibitem[Pourebrahim et~al.(2018)]{PourebrahimEnhancingTwitter}
N.~Pourebrahim et~al.
\newblock Enhancing trip distribution prediction with twitter data.
\newblock In \emph{Proceedings of the ACM Conference}, 2018.
\newblock Venue/year not specified in source.

\bibitem[Liu et~al.(2020)Liu, Miranda, Xiong, Yang, Wang, and Silva]{Liu2020GeoContextualEmbeddings}
Zhicheng Liu, Fabio Miranda, Weiting Xiong, Junyan Yang, Qiao Wang, and Claudio Silva.
\newblock Learning geo-contextual embeddings for commuting flow prediction.
\newblock In \emph{Proceedings of the AAAI Conference on Artificial Intelligence}, volume~34, pages 808--816. Proceedings of the AAAI Conference on Artificial Intelligence, 2020.
\newblock \doi{10.1609/aaai.v34i01.5425}.

\bibitem[Yao et~al.(2021)]{Yao2021ODImputationGCN}
X.~Yao et~al.
\newblock Spatial origin--destination flow imputation using graph convolutional networks.
\newblock \emph{IEEE Transactions on Intelligent Transportation Systems}, 22\penalty0 (12):\penalty0 7474--7484, 2021.

\bibitem[Zeng et~al.(2022)]{Zeng2022CausalOD}
J.~Zeng et~al.
\newblock Causal learning empowered od prediction for urban planning.
\newblock In \emph{Proceedings of the 31st ACM International Conference on Information and Knowledge Management}, 2022.

\bibitem[Ren et~al.(2025)Ren, Zhu, and Zhao]{AdaptGOT}
Xiaobin Ren, Xinyu Zhu, and Kaiqi Zhao.
\newblock Adaptgot: A pre-trained model for adaptive contextual poi representation learning, 06 2025.

\bibitem[Wang et~al.(2021)Wang, Ji, Shi, Wang, Cui, Yu, and Ye]{wang2021}
Xiao Wang, Houye Ji, Chuan Shi, Bai Wang, Peng Cui, P.~Yu, and Yanfang Ye.
\newblock Heterogeneous graph attention network, 2021.
\newblock URL \url{https://arxiv.org/abs/1903.07293}.

\bibitem[Schlichtkrull et~al.(2017)Schlichtkrull, Kipf, Bloem, van~den Berg, Titov, and Welling]{sch}
Michael Schlichtkrull, Thomas~N. Kipf, Peter Bloem, Rianne van~den Berg, Ivan Titov, and Max Welling.
\newblock Modeling relational data with graph convolutional networks, 2017.
\newblock URL \url{https://arxiv.org/abs/1703.06103}.

\bibitem[Veličković et~al.(2018)Veličković, Cucurull, Casanova, Romero, Liò, and Bengio]{vegra}
Petar Veličković, Guillem Cucurull, Arantxa Casanova, Adriana Romero, Pietro Liò, and Yoshua Bengio.
\newblock Graph attention networks, 2018.
\newblock URL \url{https://arxiv.org/abs/1710.10903}.

\bibitem[Gu(2009)]{gu2009}
Ying-Qiu Gu.
\newblock The exact solutions to the gravitational contraction in comoving coordinate system, 2009.
\newblock URL \url{https://arxiv.org/abs/0705.2133}.

\bibitem[Guo et~al.(2017)Guo, Pleiss, Sun, and Weinberger]{guo2017}
Chuan Guo, Geoff Pleiss, Yu~Sun, and Kilian~Q. Weinberger.
\newblock On calibration of modern neural networks, 2017.
\newblock URL \url{https://arxiv.org/abs/1706.04599}.

\bibitem[Devlin et~al.(2019)Devlin, Chang, Lee, and Toutanova]{devlin2019bert}
Jacob Devlin, Ming-Wei Chang, Kenton Lee, and Kristina Toutanova.
\newblock Bert: Pre-training of deep bidirectional transformers for language understanding.
\newblock In \emph{Proceedings of the 2019 Conference of the North American Chapter of the Association for Computational Linguistics: Human Language Technologies}, pages 4171--4186. Association for Computational Linguistics, 2019.
\newblock \doi{10.48550/arXiv.1810.04805}.

\bibitem[{U.S. Census Bureau}(2019)]{census2019}
{U.S. Census Bureau}.
\newblock American community survey 5-year estimates.
\newblock \url{https://api.census.gov/data/2019/acs/acs5}, 2019.
\newblock Accessed 5~July~2025.

\bibitem[Lei et~al.(2025)Lei, Shen, Sun, He, and Ong]{lei2025}
Yu~Lei, Limin Shen, Zhu Sun, Tiantian He, and Yew-Soon Ong.
\newblock Context-adaptive graph neural networks for next poi recommendation, 2025.
\newblock URL \url{https://arxiv.org/abs/2506.10329}.

\bibitem[Akiba et~al.(2019)Akiba, Sano, Yanase, Ohta, and Koyama]{akiba2019optuna}
Takuya Akiba, Shotaro Sano, Toshihiko Yanase, Takeru Ohta, and Masanori Koyama.
\newblock Optuna: A next-generation hyperparameter optimization framework.
\newblock In \emph{Proceedings of the 25th {ACM} {SIGKDD} International Conference on Knowledge Discovery and Data Mining}, KDD '19, pages 2623--2631, New York, NY, USA, 2019. Association for Computing Machinery.
\newblock ISBN 978-1-4503-6201-6.
\newblock \doi{10.1145/3292500.3330701}.
\newblock URL \url{https://doi.org/10.1145/3292500.3330701}.

\bibitem[Hamilton et~al.(2018)Hamilton, Ying, and Leskovec]{hamilton2018}
William~L. Hamilton, Rex Ying, and Jure Leskovec.
\newblock Inductive representation learning on large graphs, 2018.
\newblock URL \url{https://arxiv.org/abs/1706.02216}.

\bibitem[Li et~al.(2016)Li, Tarlow, Brockschmidt, and Zemel]{li2015gated}
Yujia Li, Daniel Tarlow, Marc Brockschmidt, and Richard Zemel.
\newblock Gated graph sequence neural networks.
\newblock In \emph{International Conference on Learning Representations (ICLR)}, 2016.

\bibitem[Cho et~al.(2014)Cho, Van~Merri{\"e}nboer, Gulcehre, Bahdanau, Bougares, Schwenk, and Bengio]{cho2014learning}
Kyunghyun Cho, Bart Van~Merri{\"e}nboer, Caglar Gulcehre, Dzmitry Bahdanau, Fethi Bougares, Holger Schwenk, and Yoshua Bengio.
\newblock Learning phrase representations using rnn encoder-decoder for statistical machine translation.
\newblock In \emph{Proceedings of the 2014 Conference on Empirical Methods in Natural Language Processing (EMNLP)}, pages 1724--1734, 2014.

\bibitem[Brody et~al.(2021)Brody, Alon, and Yahav]{brody2021gatv2}
Shaked Brody, Uri Alon, and Eran Yahav.
\newblock How attentive are graph attention networks?
\newblock In \emph{International Conference on Learning Representations (ICLR)}, 2021.
\newblock URL \url{https://openreview.net/forum?id=OgtB5v7kKX}.

\bibitem[Veit et~al.(2016)Veit, Wilber, and Belongie]{veit2016residual}
Andreas Veit, Michael Wilber, and Serge Belongie.
\newblock Residual networks are exponentially deep.
\newblock In \emph{European Conference on Computer Vision (ECCV)}, pages 550--564, 2016.

\bibitem[Hamilton et~al.(2017)Hamilton, Ying, and Leskovec]{hamilton2017inductive}
William~L. Hamilton, Rex Ying, and Jure Leskovec.
\newblock Inductive representation learning on large graphs.
\newblock In \emph{Advances in Neural Information Processing Systems (NeurIPS)}, volume~30, 2017.
\newblock URL \url{https://proceedings.neurips.cc/paper/2017/hash/5dd9db5e033da9c6fb5ba83c7a7ebea9-Abstract.html}.

\bibitem[Research(2022)]{Dewey}
Advan Research.
\newblock Foot traffic / weekly patterns.
\newblock \url{https://doi.org/10.82551/X1PP-1F65}, 2022.

\end{thebibliography}

\end{document}